# Optical spectral asymmetry enables differential refractive-index sensing under near-perfect common-mode suppression in a single-cavity dual-comb fiber laser

Masayuki Higaki[1,†], Yoshiaki Nakajima[2,†], Shuji Taue[3], Eiji Hase[4,5], Yu Tokizane[4,5], and Takeshi Yasui[4,5,*]

[1]Graduate School of Sciences and Technology for Innovation, Tokushima University, 2-1, Minami-Josanjima, Tokushima, Tokushima 770-8506, Japan

[2]Department of Physics, Toho University, 2-2-1 Miyama, Funabashi, Chiba 274-8510, Japan

[3]School of System Engineering, Kochi University of Technology, 185, Miyanokuchi, Tosayamada, Kami, Kochi 782-8502, Japan

[4]Institute of Post-LED Photonics (pLED), Tokushima University, 2-1, Minami-Josanjima, Tokushima, Tokushima 770-8506, Japan

[5]Institute of Photonics and Human Health Frontier (IPHF), Tokushima University, 2-1, Minami-Josanjima, Tokushima, Tokushima 770-8506, Japan

[†]These authors contributed equally.

**Keywords:** Optical frequency comb / Dual-comb / Refractive-index sensing / Fiber laser / Optical spectral asymmetry / Common-mode suppression

*Takeshi Yasui, E-mail: yasui.takeshi@tokushima-u.ac.jp

## Abstract

Single-cavity dual-comb fiber lasers offer near-perfect common-mode suppression, but such suppression is generally expected to eliminate differential sensing signals. Here we show that optical spectral asymmetry enables differential refractive-index (RI) sensing despite near-perfect common-mode suppression. By integrating an intracavity multimode interference (MMI) fiber sensor into a shared cavity, two counter-propagating combs are simultaneously generated and interrogated, enabling differential RI sensing based on the repetition-frequency difference ($\Delta f_{rep}$). This architecture achieves near-perfect common-mode suppression, reducing temperature-induced drift by more than two orders of magnitude compared with a conventional dual-cavity dual-comb system. Spectral analysis reveals that optical spectral asymmetry gives rise to asymmetric repetition-frequency responses through cavity dispersion, thereby providing the physical origin of the differential sensing signal. A trade-off between sensitivity and noise suppression is identified: while the sensitivity is reduced due to the shared sensor, the improved frequency stability enhances robustness against environmental perturbations. These results demonstrate that near-perfect common-mode suppression does not preclude differential sensing, but instead reveals a new sensing mechanism governed by optical spectral asymmetry. These findings establish optical spectral asymmetry as a design principle for differential sensing in single-cavity dual-comb systems and provide a framework for highly stable long-term photonic sensing.

## 1. Introduction

Refractive index (RI) sensing is a fundamental technique widely used in material characterization [1] and biomedical applications [2]. Recent demands require the simultaneous realization of high precision, long-term stability, and fast response. Optical frequency comb (OFC)–based RI sensing [3-5], hereafter referred to as an RI-sensing OFC, has emerged as a promising approach, in which RI-induced changes are encoded into shifts in the repetition frequency ($f_{rep}$). By utilizing the unique frequency-linking capability of OFCs, this scheme enables high-precision readout in the radio-frequency (RF) domain by linking optical sensing to electrical detection. However, measurement stability is often limited by environmental perturbations, particularly temperature-induced drift, which introduces fluctuations in $f_{rep}$ and degrades performance.

Dual-comb configurations have emerged as an effective approach to improve the stability of RI sensing [6,7]. In particular, active–dummy architectures enable suppression of temperature-induced drift through common-mode rejection. By monitoring the differential repetition frequency ($\Delta f_{rep} = f_{rep1} - f_{rep2}$) between the two combs, environmental fluctuations can be significantly reduced. In mechanically shared dual-cavity configurations [8,9], both combs experience largely similar thermal perturbations, resulting in a high degree of correlation in their responses. As a consequence, long-term stability and measurement reproducibility are substantially improved, making dual-comb differential RI sensing a state-of-the-art approach.

However, the degree of common-mode suppression in dual-cavity dual-comb configurations is fundamentally limited. In practice, the two combs do not perfectly share the same environment, resulting in incomplete correlation between their responses. Non-common-mode noise arises from unshared optical paths and components, which cannot be fully canceled in the differential measurement. Consequently, residual fluctuations remain in $\Delta f_{rep}$, limiting the achievable stability. This imperfect common-mode correlation ultimately determines the sensing performance, revealing an intrinsic limitation of dual-comb differential RI sensing systems.

A single-cavity dual-comb configuration provides a promising solution to maximize common-mode correlation. Dual OFCs can be generated from a single cavity using approaches such as wavelength multiplexing [10,11], polarization multiplexing [12,13], and bidirectional multiplexing [14-16]. Because both combs share the same optical cavity, they experience highly similar environmental perturbations, leading to a significantly higher degree of correlation compared to mechanically shared dual-comb systems. This configuration effectively suppresses common-mode noise, thereby reducing residual fluctuations in $\Delta f_{rep}$. The effectiveness of such single-cavity dual-comb systems has been widely demonstrated in various applications [17], including optical spectroscopy [10,18], ranging [19,20], THz spectroscopy [21,22], and asynchronous optical sampling (ASOPS) [11,23], where precise and/or rapid measurements are achieved without strict control of the individual combs. However, their application to RI sensing has remained largely unexplored, despite the potential for further improvements in stability and measurement reproducibility.

More importantly, even if near-perfect common-mode suppression can be achieved, a fundamental question remains unresolved. Near-perfect common-mode suppression would naively be expected to suppress not only environmental noise but also the differential sensing signal itself. Whether differential RI sensing can survive under such conditions therefore remains an open question. Here, we demonstrate that optical spectral asymmetry enables differential RI sensing in a bidirectional multiplexing-based single-cavity dual-comb system despite near-perfect common-mode suppression. The physical concept and operating principle of the proposed scheme are schematically illustrated in Fig. 1, which compares the conventional dual-cavity and the proposed single-cavity configurations. In the former, temperature drift is only quasi-common-mode because the active and dummy combs propagate in separate cavities. In contrast, the shared intracavity sensor and cavity in the proposed configuration enable near-perfect common-mode temperature compensation while preserving a measurable differential signal through optical spectral asymmetry between the CW and CCW combs. As will be shown, the proposed configuration simultaneously achieves more than two orders of magnitude suppression of temperature-induced drift, thereby establishing optical spectral asymmetry as a design principle for differential sensing in single-cavity dual-comb systems.

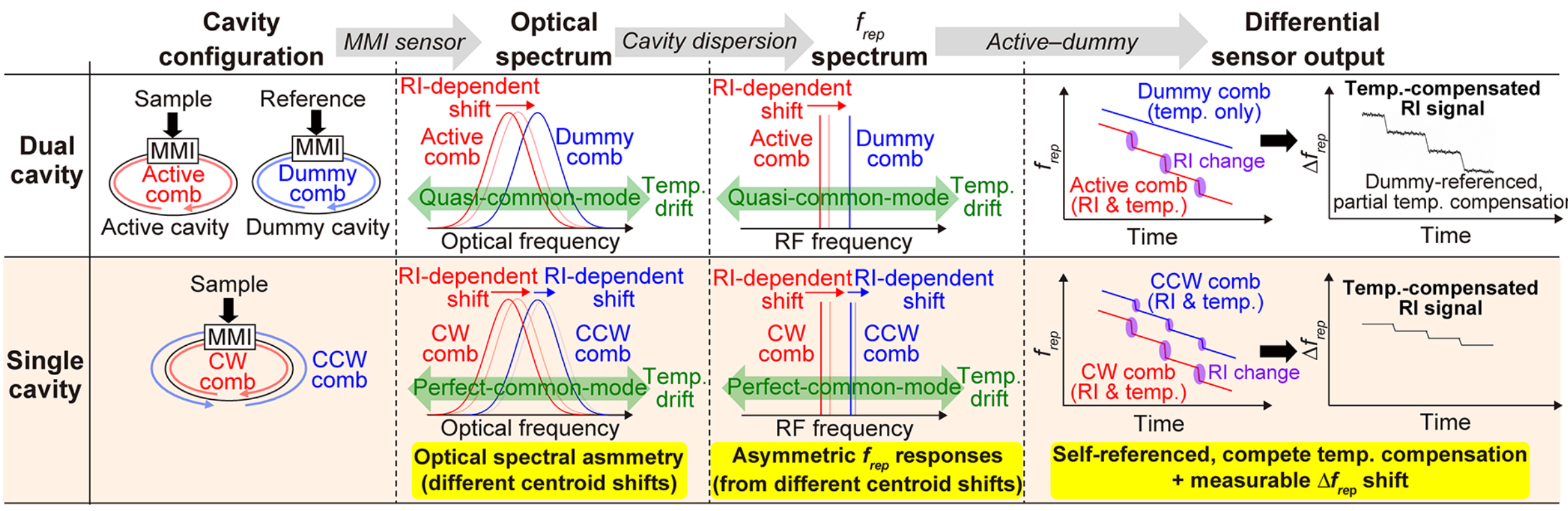

Fig. 1. Conceptual comparison of conventional dual-cavity and proposed single-cavity dual-comb refractive-index (RI) sensing. In the dual-cavity configuration, temperature fluctuations are only quasi-common-mode, resulting in partial cancellation by dummy-referenced compensation. In the proposed single-cavity configuration, the CW and CCW combs share the same intracavity MMI sensor and cavity, enabling nearly perfect common-mode temperature compensation. Despite the shared sensing architecture, optical spectral asymmetry between the counter-propagating combs results in asymmetric repetition-frequency responses through cavity dispersion, thereby preserving a measurable differential signal ($\Delta f_{rep}$ shift).

## 2. Methods

The conceptual mechanism of the proposed approach is illustrated in Fig. 1, whereas the detailed experimental implementation is shown in Fig. 2. The detailed measurement principle of RI-sensing OFC has been described in our previous studies [3–7]; therefore, only a brief overview is provided here. In RI-sensing OFCs, RI changes are converted into RI-dependent optical spectral shifts, as the intracavity MMI fiber sensor functions as an RI-dependent tunable optical bandpass filter. These spectral shifts are then translated into changes in the effective optical cavity length ($nL$), where $n$ and $L$ denote the fiber refractive index and geometric length, respectively, through the dispersion of the cavity fiber. As a result, RI changes are finally converted into shifts in $f_{rep}$, since $f_{rep} = c/(nL)$, where $c$ is the speed of light in vacuum. As schematically illustrated in the upper panel of Fig. 1, dual-comb active–dummy temperature compensation employs a pair of RI-sensing OFCs with nominally identical characteristics. One comb functions as an active OFC ($f_{rep1}$), which is sensitive to both sample RI and environmental temperature, while the other comb serves as a dummy

OFC ($f_{rep2}$), which is sensitive only to environmental temperature. Because environmental temperature fluctuations are only quasi-common-mode between the two OFCs, the differential repetition frequency ($\Delta f_{rep} = f_{rep1} - f_{rep2}$) provides only partial cancellation of temperature-induced drift, thereby enabling RI measurement in the presence of residual drift.

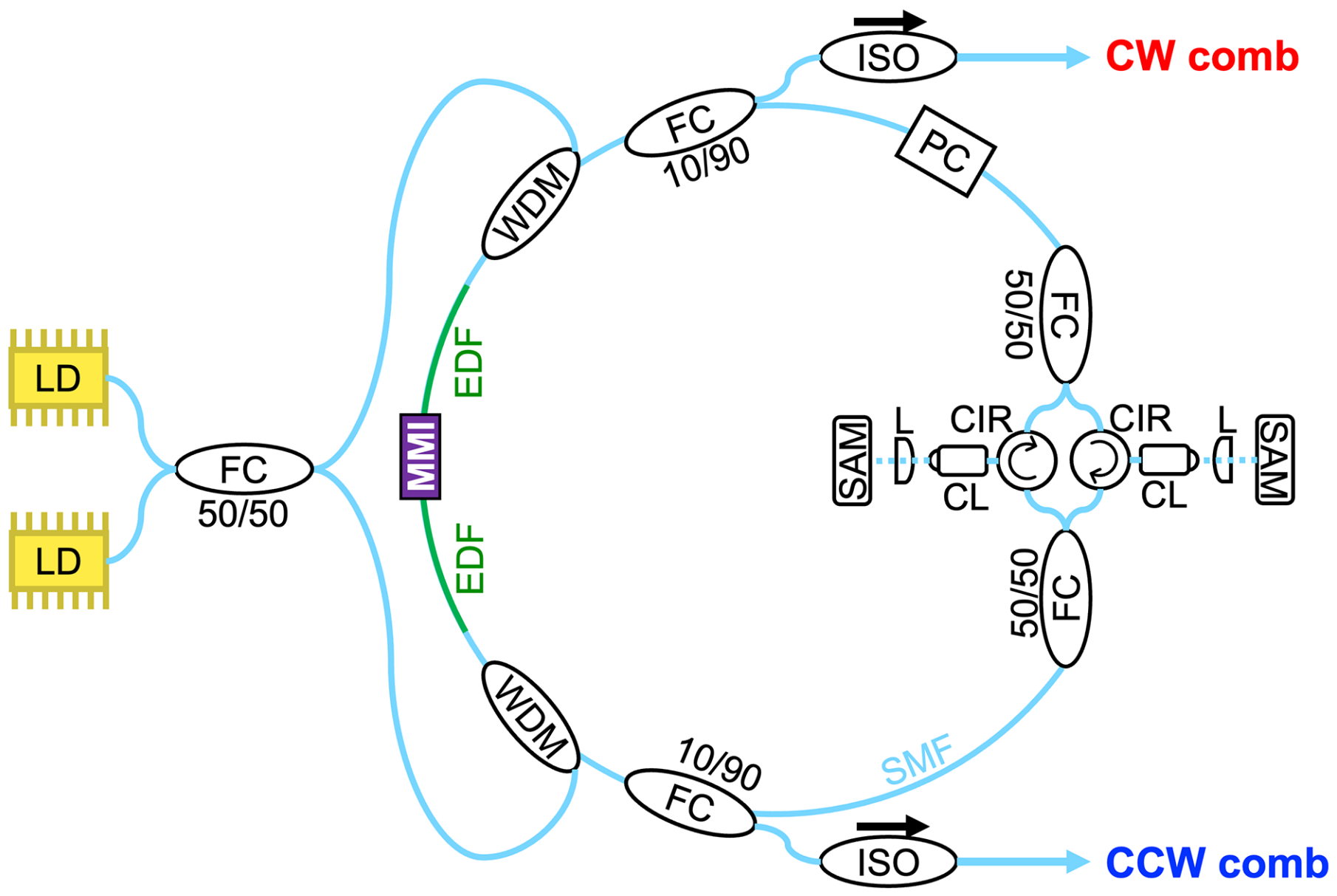

Fig. 2. Schematic drawing of experimental setup. MMI, intracavity multimode interference fiber sensor; EDF, erbium-doped single-mode fiber; WDMs, wavelength-division multiplexers; FCs, fiber couplers; PC, polarization controller; LDs, pump laser diodes; ISOs, isolators; CIRs, circulators; CLs, collimators; Ls, lensed; SAMs, saturable absorber mirrors.

Among various single-cavity dual-comb configurations, bidirectional multiplexing is particularly well suited to MMI-based sensing, as it avoids polarization and wavelength mismatches and preserves a nearly perfect common-mode response to environmental perturbations. In this configuration, an MMI fiber sensor is inserted into the shared cavity and is simultaneously interrogated by two counter-propagating combs (Fig. 1, lower panel). Although the same sensor is shared, optical spectral asymmetry between the counter-propagating combs results in slightly different repetition-frequency responses through cavity dispersion, thereby producing a measurable RI-dependent $\Delta f_{rep}$ shift. Integrating an MMI sensor into the shared cavity can be challenging, as it introduces optical loss, spectral filtering, and polarization scrambling. Nevertheless, by optimizing the cavity parameters, stable bidirectional dual-comb operation is achieved within a single cavity.

Figure 2 shows the experimental setup of the single-cavity dual-comb differential RI sensing system based on bidirectional multiplexing. The cavity design is based on a previously reported ring-type fiber laser configuration [15]. A key aspect is the placement of the intracavity MMI fiber sensor (MMI wavelength = 1556.6 nm), which is composed of a clad-less multimode fiber (core diameter = 125 μm, length = 58.94 mm) spliced between two single-mode fibers (core diameter = 8.5 μm). The MMI sensor is positioned between erbium-doped fiber (EDF) sections to compensate for optical loss and spectral filtering induced by the MMI, while preserving balanced bidirectional gain conditions required for stable broadband dual-comb mode locking. The shared cavity consists of single-mode fiber (SMF, SMF28e, Corning; dispersion at 1550 nm = 18 ps·km$^{-1}$·nm$^{-1}$), EDF(EDF-80, OFS; dispersion at 1550 nm = -40 ps·km$^{-1}$·nm$^{-1}$; core absorption = 80 dB/m at 1530 nm), fiber couplers (FCs), wavelength-division multiplexers (WDMs), a polarization controller (PC), and the MMI sensor, while circulators (CIRs), collimators (CLs), lenses (Ls), saturable absorbers (SAMs), and isolators (IOSs) are placed in the non-shared sections. Despite these non-shared elements, more than 77% of the cavity is shared. To compensate for the relatively large loss introduced by the intracavity MMI sensor, pumping using two laser diodes (LDs; wavelength = 976 nm; optical power = 950 mW each) was employed to ensure sufficient intracavity gain for stable mode-locked operation. Mode locking was achieved using a combination of nonlinear polarization rotation and saturable absorption. While nonlinear polarization rotation is controlled by a common polarization controller acting on both combs, independent SAMs are employed for the two counter-propagating combs, enabling stable dual-comb operation without mutual competition within a single cavity. Achieving stable bidirectional mode locking with broadband spectra in the shared-cavity configuration required careful optimization of the cavity parameters and mode-locking conditions, resulting in repetition frequencies of 36.182 and 36.280 MHz for the CW and CCW combs ($f_{rep_CW}$ and $f_{rep_CCW}$), respectively.

## 3. Results

We first compare the optical spectra of the counter-propagating combs, as shown in Fig. 3. The clockwise (CW) comb exhibits an average output power of 4.86 mW (red trace), whereas the counter-clockwise (CCW) comb shows 2.36 mW (blue trace). Although the MMI fiber sensor is shared and, in principle, acts as an RI-dependent optical bandpass filter with symmetric characteristics for both propagation directions, the measured spectra exhibit noticeable differences in shape, consistent with the optical spectral asymmetry schematically illustrated in Fig. 1. This deviation is attributed to asymmetric gain dynamics associated with the EDF sections placed on both sides of the MMI sensor. In addition to compensating for the MMI-induced loss, the EDF provides optical gain and induces spectral broadening after the MMI section. The extent of these effects differs between the two propagation directions, likely due to polarization-dependent gain and propagation effects, resulting in distinct spectral profiles for the CW and CCW combs.

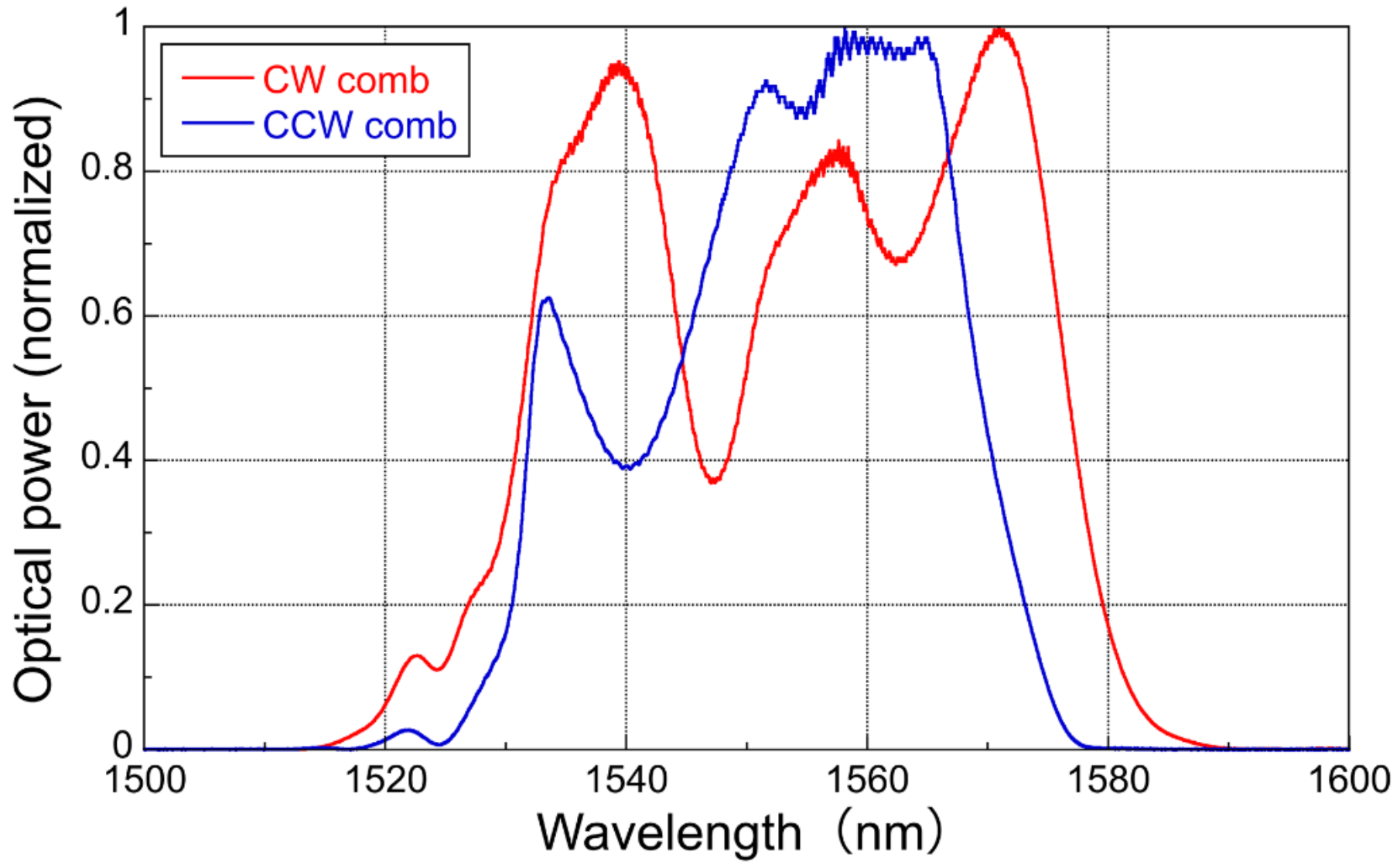

Fig. 3. Optical spectra of the counter-propagating combs. The CW (red trace, 4.86 mW, $f_{rep_CW}$ = 36.182 MHz) and CCW (blue trace, 2.36 mW, $f_{rep_CCW}$ = 36.280 MHz) combs exhibit distinct spectral profiles despite sharing the same intracavity MMI fiber sensor. This spectral asymmetry originates from direction-dependent gain and propagation effects in the cavity.

Next, we evaluate the repetition-frequency ($f_{rep}$) and differential-frequency ($\Delta f_{rep}$) characteristics. Figure 4(a) shows the temporal deviations of $f_{rep_CW}$ (red trace), $f_{rep_CCW}$ (blue trace), and their difference $\Delta f_{rep} = f_{rep_CW} - f_{rep_CCW}$ (green trace) from their initial values, measured with a gate time of 1 s and a sampling interval of 1 s. The temperature-induced drift observed in $f_{rep_CW}$ and $f_{rep_CCW}$ is highly correlated, indicating a nearly perfect common-mode response to environmental perturbations. As a result, the differential signal $\Delta f_{rep}$ exhibits a dramatic reduction in frequency fluctuations, with a standard deviation of 0.162 Hz. For comparison, the same evaluation is performed for a dual-cavity dual-comb system reported previously [6,7]. Figure 4(b) shows the repetition frequencies of the active comb ($f_{rep_act}$, red trace), the dummy comb ($f_{rep_dum}$, blue trace), and their difference $\Delta f_{rep}$ (green trace). Although $f_{rep_act}$ and $f_{rep_dum}$ exhibit generally similar trends, their responses are not perfectly matched, resulting in residual temperature drift in $\Delta f_{rep}$ with a standard deviation of 4.01 Hz. Comparison between Figs. 4(a) and 4(b) clearly indicates that the drift level in the single-cavity configuration is reduced by more than two orders of magnitude compared to the dual-cavity configuration. To further quantify the stability over different time scales, the Allan deviation of $\Delta f_{rep}$ is compared for both systems, as shown in Fig. 4(c). The single-cavity dual-comb system (red trace) consistently exhibits lower Allan deviation than the dual-cavity system (blue trace). In particular, for averaging times longer than 0.1 s, the dual-cavity system shows an increasing trend due to residual drift, whereas the single-cavity system exhibits a plateau, confirming the effectiveness of the self-referenced temperature compensation enabled by the shared-cavity configuration.

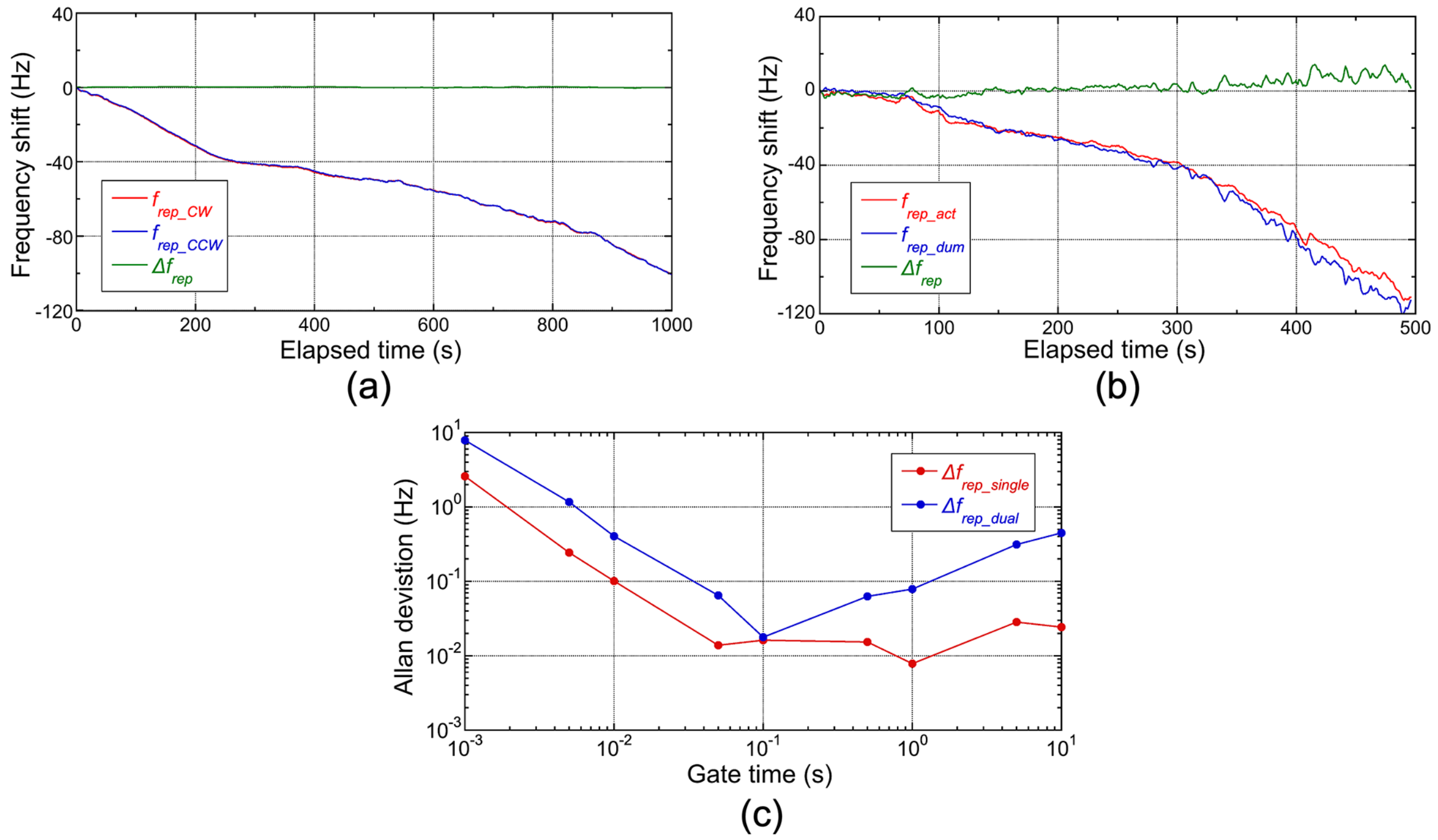

Fig. 4. Repetition-frequency stability of the dual-comb systems. (a) Temporal frequency deviations from the initial values of the repetition frequencies ($f_{rep_CW}$ and $f_{rep_CCW}$) for the clockwise (CW, red trace) and counter-clockwise (CCW, blue trace) combs, and their difference $\Delta f_{rep}$ (green trace) in the single-cavity configuration. The differential signal shows a significant reduction in frequency fluctuations due to strong common-mode correlation. (b) Corresponding results for a dual-cavity dual-comb system, showing the temporal frequency deviations from the initial values of the repetition frequencies ($f_{rep_act}$ and $f_{rep_dum}$) for the active (red trace) and dummy (blue trace) RI-sensing OFCs and their difference $\Delta f_{rep}$ (green trace), where residual drift remains. (c) Allan deviation of $\Delta f_{rep}$ for the single-cavity (red trace) and dual-cavity (blue trace) systems, demonstrating improved frequency stability in the single-cavity configuration, particularly at longer averaging times.

Finally, the RI sensing performance of the proposed system is quantitatively evaluated. The assessment is based on key performance metrics, including sensitivity, linearity, and resolution, derived from the RI dependence of $\Delta f_{rep}$. Ethanol–water mixtures with ethanol concentrations of 0%, 10%, 20%, and 30% are used as test samples, corresponding to RI values of 1.3180 RIU, 1.3213 RIU, 1.3248 RIU, and 1.3281 RIU, respectively. The RI values are calculated from reported RIs of pure water and ethanol based on their mass fractions. The uncertainty in the assigned RI values is estimated to be on the order of $10^{-5}$ RIU, which is sufficiently smaller than the RI variations evaluated in this study. Five repeated measurements are performed for each sample. Sensitivity is obtained from the slope of the linear fit between RI and the averaged frequency shift, while linearity is evaluated using the coefficient of determination ($R^2$). The resolution is defined as the ratio of the average standard deviation of repeated measurements (excluding pure water) to the sensitivity.

Figure 5(a) shows the RI sensing response of the single-cavity dual-comb configuration. As anticipated from the conceptual mechanism illustrated in Fig. 1, the frequency shift in $\Delta f_{rep}$ is relatively small due to the strong common-mode suppression in the shared MMI sensor. Nevertheless, a clear linear relationship is observed between the sample RI and the $\Delta f_{rep}$ shift, demonstrating the feasibility of RI sensing using the proposed method. To evaluate the RI sensing precision, the frequency fluctuations at each RI were converted into equivalent RI fluctuations using the linear calibration relationship in Fig. 5(a). The resulting standard deviations of the estimated RI are summarized in Fig. 5(b). The standard deviations remain nearly constant over the investigated RI range, indicating that the sensing precision is essentially independent of the sample RI. The corresponding performance metrics are as follows: the sensitivity is 737 Hz/RIU, the linearity is $R^2$ = 0.993, and the resolution is $1.32 \times 10^{-4}$ RIU. For comparison, Fig. 5(c) shows the RI sensing response obtained using the dual-cavity dual-comb configuration. A similarly strong linear dependence is observed between the sample RI and $\Delta f_{rep}$ shift; however, the individual data points exhibit noticeably larger scatter, reflecting residual temperature fluctuations that remain after dummy-referenced compensation. Using the linear calibration relationship in Fig. 5(c), the frequency fluctuations were likewise converted into equivalent RI fluctuations, and the resulting standard deviations of the estimated RI are summarized in Fig. 5(d). The corresponding performance metrics are as follows: the sensitivity is $1.87 \times 10^4$ Hz/RIU, the linearity is $R^2$ = 0.997, and the resolution is $2.09 \times 10^{-4}$ RIU. These results indicate that, although the single-cavity configuration exhibits lower sensitivity due to enhanced common-mode suppression, it maintains excellent linearity while providing improved robustness against environmental fluctuations. The reduction in sensitivity originates from the shared sensing configuration, which suppresses not only common-mode noise but also a portion of the differential sensing signal. This trade-off is intrinsic to single-cavity implementations. Nevertheless, the greatly improved temperature stability enables reliable RI sensing with reduced susceptibility to environmental perturbations.

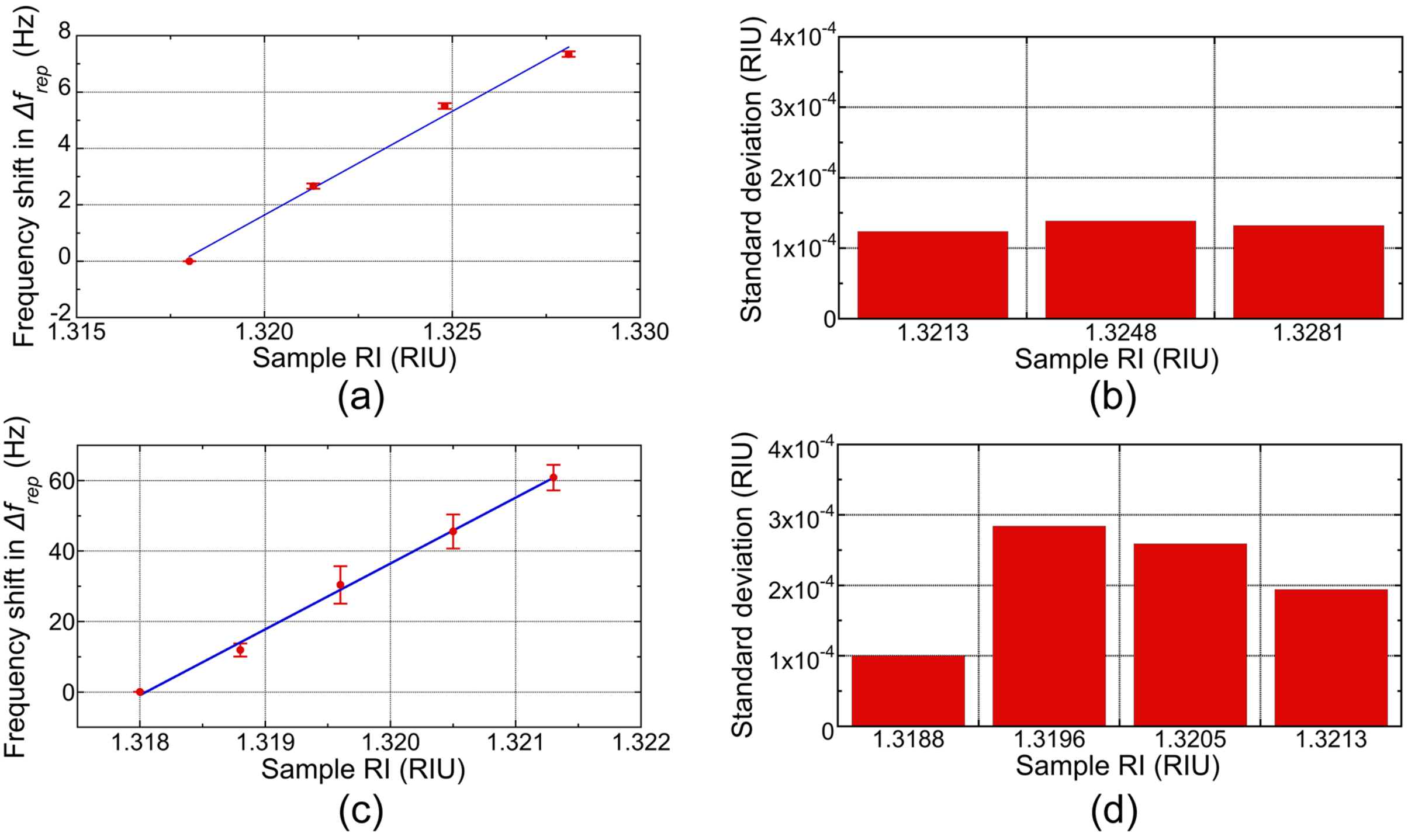

Fig. 5. Refractive-index (RI) sensing performance based on the differential repetition frequency ($\Delta f_{rep}$). (a) RI dependence of $\Delta f_{rep}$ shift in the single-cavity dual-comb configuration, showing a clear linear relationship despite reduced frequency shift due to strong common-mode suppression. Error bars indicate the standard deviation of five repeated measurements, and the solid line represents a linear fit. (b) Standard deviation of the estimated RI values obtained from the linear calibration relationship in (a). The RI fluctuations are nearly independent of the sample RI. (c) RI dependence of $\Delta f_{rep}$ shift in the dual-cavity dual-comb configuration. A larger frequency shift is observed, although the individual data points exhibit larger scatter due to residual temperature fluctuations. (d) Standard deviation of the estimated RI values obtained from the linear calibration relationship in (c).

## 4. Discussion

An intriguing aspect of the present single-cavity dual-comb configuration is that distinct RI responses are observed for the two counter-propagating combs, despite sharing the same intracavity MMI fiber sensor. This behavior is physically nontrivial and warrants further discussion. A key factor is the difference in optical spectral characteristics between the two combs, as shown in Fig. 3. Such spectral asymmetry can lead to different effective responses of the MMI sensor, even though its structural properties are symmetric. To quantitatively evaluate the spectral asymmetry, the RI-dependent spectral evolution of each comb was analyzed. Figures 6(a) and 6(b) show the RI-dependent optical spectra of the CW and CCW combs, respectively. In both combs, clear RI-dependent spectral shifts are observed. To quantitatively evaluate these spectral shifts, the spectral centroid wavelength was calculated for each measured spectrum. The centroid wavelength was defined as the intensity-weighted average wavelength of the optical spectrum. Figure 6(c) shows the RI dependence of the centroid wavelengths for the CW (red trace) and CCW (blue trace) combs. Both combs exhibit clear RI-dependent shifts, confirming that the MMI sensor induces spectral modulation in response to RI changes. Linear fitting yields sensitivities of 84.7 nm/RIU for the CW comb and 73.9 nm/RIU for the CCW comb, indicating a noticeable difference in spectral response between the two propagation directions. In RI-sensing OFCs, RI-induced spectral shifts are converted into $f_{rep}$ shifts through cavity dispersion. To quantitatively estimate this conversion process, the dispersion characteristics of the cavity fiber were analyzed. The total group-delay dispersion (GDD) of the cavity was estimated to be 0.068452 $ps^2$, corresponding to a wavelength dependence of the cavity round-trip time ($dT/d\lambda$) of $-5.36524 \times 10^{-2}$ ps/nm. From this value, the wavelength dependence of the repetition frequency was separately calculated for the CW and CCW propagating combs. The obtained values were dfrep/dλ = 70.24 Hz/nm for the CW comb and 70.62 Hz/nm for the CCW comb. Combining the calculated $df_{rep}/d\lambda$ with the experimentally obtained centroid sensitivities yields the RI dependences of the repetition-frequency shifts for the CW and CCW combs, as shown in Fig. 6(d). Linear fitting of the calculated responses yields estimated $f_{rep}$ sensitivities of 5946.1 and 5218.3 Hz/RIU for the CW and CCW combs, respectively, corresponding to a differential sensitivity of 727.8 Hz/RIU. For the RI variation used in the experiment ($\Delta RI$ = 0.0101 RIU, corresponding to pure water and 30% ethanol solution), the expected $\Delta f_{rep}$ shift is calculated to be 7.35 Hz, which agrees well with the experimental result shown in Fig. 5(a). The difference in spectral centroid shift directly results in asymmetric $f_{rep}$ responses via cavity dispersion, which gives rise to the observed $\Delta f_{rep}$ shift. These results indicate that the asymmetric spectral response is the dominant physical origin of the differential sensing signal in the single-cavity dual-comb system. The spectral asymmetry can be controlled through cavity design, suggesting a route to engineer $\Delta f_{rep}$ sensitivity in single-cavity systems. Importantly, these results reveal that highly correlated common-mode suppression does not eliminate sensing capability, but instead enables a new sensing mechanism based on optical spectral asymmetry between the counter-propagating combs.

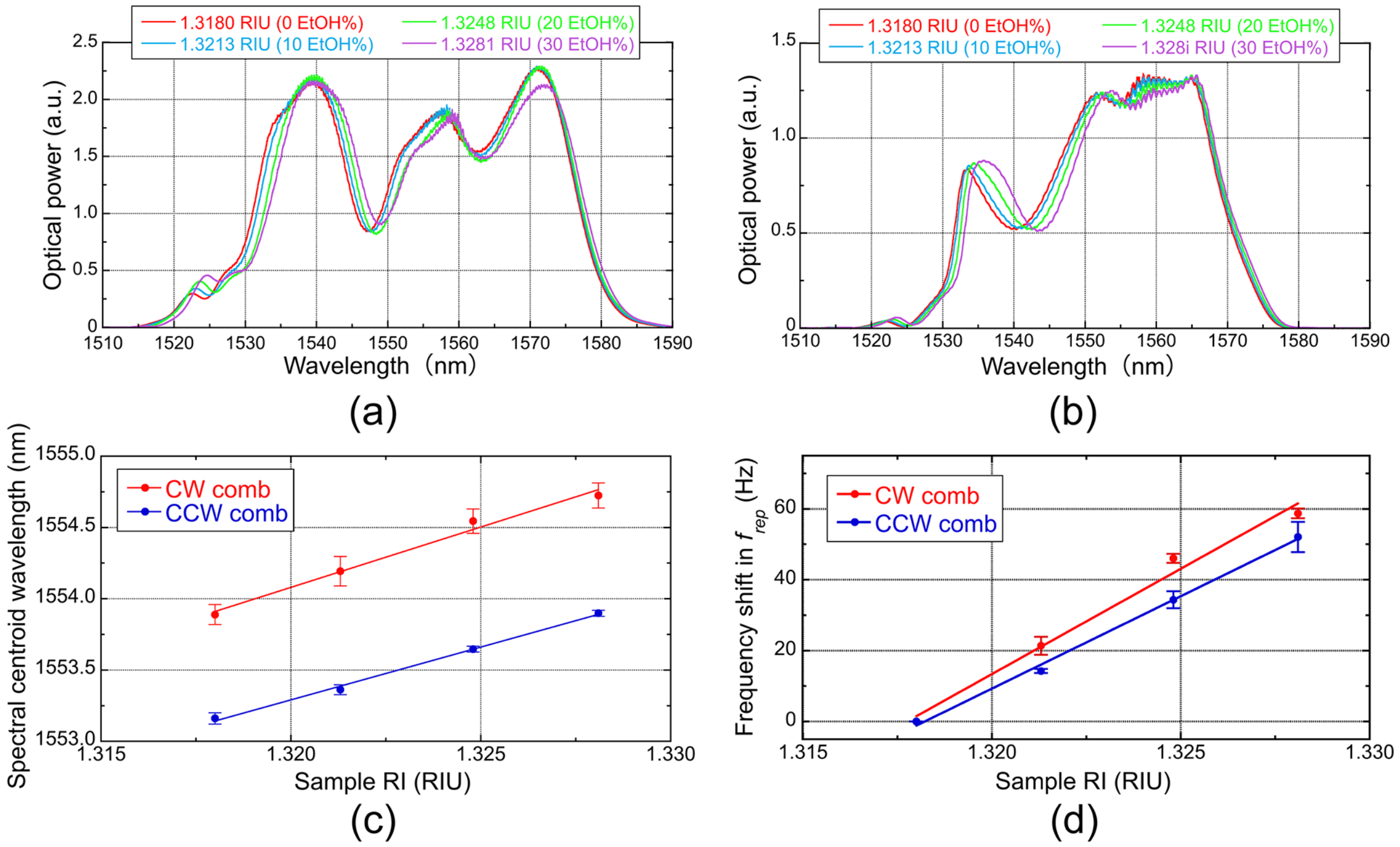

Fig. 6. RI-dependent optical spectral characteristics of the counter-propagating combs in the single-cavity dual-comb system. (a) Optical spectra of the CW comb measured at different sample RIs. (b) Optical spectra of the CCW comb measured at different sample RIs. (c) RI dependence of the spectral centroid wavelength for the CW (red trace) and CCW (blue trace) combs. Solid lines represent linear fitting results, yielding sensitivities of 84.7 and 73.9 nm/RIU for the CW and CCW combs, respectively. (d) Calculated RI dependence of the repetition-frequency shifts for the CW (red trace) and CCW (blue trace) combs obtained from the measured centroid shifts and the cavity dispersion. Solid lines represent linear fitting results, yielding estimated $f_{rep}$ sensitivities

of 5946.1 and 5218.2 Hz/RIU for the CW and CCW combs, respectively.

While the preceding analysis clarifies the physical origin of the sensing signal, an important trade-off between sensitivity and noise suppression is observed in the present single-cavity configuration. Compared to the dual-cavity system, the sensitivity is reduced by approximately one order of magnitude, whereas the temperature-induced drift is suppressed by more than two orders of magnitude. This behavior can be understood as a consequence of the shared sensing configuration, which suppresses not only common-mode noise but also differential signal components. This result highlights an important feature of single-cavity dual-comb systems, namely the orthogonal relationship between signal scaling and noise suppression. While the signal amplitude is reduced due to the shared sensor, the substantial improvement in frequency stability leads to enhanced robustness against environmental perturbations. As a result, the overall sensing performance depends on the measurement time scale and environmental conditions. In the present RI sensing experiment, the advantage of drift suppression is not fully reflected in the RI resolution due to the relatively short measurement duration and moderate environmental fluctuations. However, for long-term monitoring applications, where temperature drift is a dominant limiting factor, the improved stability of the single-cavity configuration is expected to provide significant benefits. Furthermore, the reduced sensitivity could be compensated by combining this approach with signal-enhancement techniques, such as dual-terahertz-comb-based schemes [24], which enable independent scaling of the sensing signal without compromising noise suppression. This trade-off highlights a new design paradigm for dual-comb sensing systems, where sensitivity and stability can be independently optimized depending on the target application.

## 5. Conclusions

In this study, we demonstrated RI sensing using a bidirectional multiplexing-based single-cavity dual-comb system. By integrating an intracavity MMI fiber sensor into a shared cavity, two counter-propagating combs were simultaneously generated and interrogated, enabling differential RI sensing based on $\Delta f_{rep}$. The proposed configuration achieved a substantial reduction in temperature-induced drift, exceeding two orders of magnitude compared to a conventional dual-cavity system. Despite the shared sensing configuration and the resulting strong common-mode suppression, a measurable $\Delta f_{rep}$ response was preserved owing to the intrinsic optical spectral asymmetry between the counter-propagating combs. This finding clarifies that the differential sensing signal is governed by asymmetric $f_{rep}$ responses induced through cavity dispersion. These results reveal that nearly perfect common-mode suppression does not eliminate sensing capability, but instead enables a new sensing mechanism based on optical spectral asymmetry. Furthermore, a trade-off between sensitivity and noise suppression was identified, highlighting an orthogonal relationship between signal scaling and stability. These findings establish optical spectral asymmetry as a design principle for differential sensing in single-cavity dual-comb systems, enabling sensitivity and robustness to be tailored according to application requirements. The proposed sensing mechanism opens new opportunities for highly stable long-term photonic sensing, including biochemical sensing, environmental monitoring, and industrial process control, where environmental drift is a dominant limiting factor.

**Data Availability**

The data that support the findings of this study are available from the corresponding author upon reasonable request.

**Funding**

This work was supported in part by the Japan Society for the Promotion of Science (JPJS00420240022, 24K21237, 26KJ1756); Japan Science and Technology Agency (JPMJSP2113).

**Conflict of Interest**

The authors have no conflicts of interest to disclose.

**Author Contributions**

**Masayuki Higaki**: Investigation (equal); Data curation (equal); Methodology (equal); Writing - review & editing (equal); Funding acquisition (supporting). **Yoshiaki**

**Nakajima:** Investigation (equal); Data curation (equal); Methodology (equal). **Shuji Taue:** Methodology (supporting). **Eiji Hase:** Supervision (supporting). **Yu Tokizane:** Supervision (supporting). **Takeshi Yasui:** Conceptualization (lead); Project administration (lead); Methodology (equal); Writing - original draft (lead); Writing - review & editing (equal); Supervision (lead); Funding acquisition (lead).